\documentclass[twocolumn,showpacs,epsf,preprintnumbers,amsmath,amssymb]{revtex4}

\usepackage{graphicx}
\usepackage{dcolumn}
\usepackage{bm}

\begin{document}
\title{  Polarization and decoherence in  a two-component Bose-Einstein Condensate}
\author{Le-Man Kuang$^{1,2}$, Jin-Hui Li$^{1}$ and Bambi Hu$^{2,3}$}
\address{$^1$Department of Physics, Hunan Normal University, Changsha 410081, China\\
 $^2$Department of Physics, University of Houston, Houston, Texas 77204\\
 $^3$Department of Physics and Centre for Nonlinear Studies, Hong Kong Baptist University,  Hong Kong, China}

\begin{abstract}
We theoretically investigate polarization properties  of a
two-component Bose-Einstein condensate (BEC) and influence of
decoherence induced by environment on BEC polarization through
introducing four BEC Stokes operators which  are quantum analog of
the classical Stokes parameters for a light field. BEC
polarization states can be geometrically described by a
Poincar\'{e} sphere defined by expectation values of BEC Stokes
operators. Without decoherence, it is shown that nonlinear
inter-atomic interactions in the BEC induce periodic polarization
oscillations whose periods depend on the difference between
self-interaction in each component and inter-component interaction
strengths. In particular,  when inter-atomic nonlinear
self-interaction in each BEC component equals inter-component
nonlinear interaction, Stokes vector associated with Stokes
operators precesses around a fixed axis in the dynamic evolution
of the BEC. The value of the processing frequency is determined by
the strength of the linear coupling between two components of the
BEC. When decoherence is involved, we find each component of the
Stokes vector decays which implies that
 decoherence depolarizes the BEC.
\end{abstract}

\maketitle
\section{Introduction}
The  experimental realization of Bose-Einstein condensation \cite{and}
 and an atom laser \cite{mew} have sparked many theoretical and experimental
 studies of coherent atomic matter.
 It is well understood that matter-wave analog of nonlinear optical interactions
 is provided by inter-atomic collisions. In particular, in the $s$-wave scattering
 regime,  two-body collisions are mathematically equivalent  to a optical Kerr
 interaction.  Hence,  many of the concepts of first developed in optics  can
 readily be extended to Bose-Einstein condensates (BEC). Recently, much attention
  has been paid to    these lines, including the study
 of matter-wave solitons \cite{mor}, phase conjugation \cite{gol}, four-wave mixing \cite{mar},
 an atom laser \cite{wis}, atom holography \cite{ozob}, etc. Especially, pulsed  and
 continuous-wave  atom lasers \cite{hag} and  the four-wave mixing with matter waves
 \cite{den} have been  demonstrated experimentally. The experimental verification
 of the other predictions is also underway.

As is well known that polarization \cite{col} is an important
property that is common to all types of vector waves. Light waves
possess this property and so do elastic and spin waves in solids,
for example. In this paper we study BEC polarization  that arises
for matter waves by considering  the two components of a
two-component BEC as two polarizations. The motivation for present
study is twofold. First, the BEC polarization  is  interesting in
itself and deserves further examination. From a basic point of
view, it significantly  extends the already well-established range
of analogies between light and matter waves.  Second, more
importantly perhaps, it may also have useful practical
applications. We shall investigate nonlinear polarization dynamics
of the BEC and study influence of  decoherence induced by
noncondensate atoms.

To begin we  consider a  two-component BEC described by the second
quantized Hamiltonian \cite{gor}
\begin{eqnarray}
\label{e1}
\hat{H}&=&\hat{H}_1+\hat{H}_2+\hat{H}_{nl} +\hat{H}_{l}, \\
\label{e2} \hat{H}_i&=&\int d{\bf x} \hat{\psi}^{\dagger}_i({\bf
x})[-\frac{\hbar^2}{2m}\nabla^2 +
V_i({\bf x})\nonumber \\
& & + U_i\hat{\psi}^{\dagger}_i({\bf x})\hat{\psi}_i({\bf
x})]\hat{\psi}_i({\bf x}), (i=1,2), \\
\label{e3}
\hat{H}_{nl}&=&U_{12} \int d{\bf x}
\hat{\psi}^{\dagger}_1({\bf x})\hat{\psi}^{\dagger}_2({\bf x})\hat{\psi}_1({\bf x})\hat{\psi}_2({\bf x}),\\
\label{e4}
 \hat{H}_{l}&=&\Lambda \int d{\bf x}
[\hat{\psi}^{\dagger}_1({\bf x})\hat{\psi}_2({\bf x}) +
\hat{\psi}_1({\bf x})\hat{\psi}^{\dagger}_2({\bf x})].
\end{eqnarray}
Here $i=1,2$, $\hat{\psi}_i({\bf x})$ and $\hat{\psi}^{\dagger}_i({\bf x})$ are the
atomic field operators which annihilate and create atoms at position ${\bf x}$, respectively.
  $\hat{H}_1$ and $\hat{H}_2$  describe the evolution of each component in the absence of
intercommunion interactions. $\hat{H}_{nl}$  is a nonlinear
interaction term, it describes inter-component collisions.
$\hat{H}_{l}$ is a linear interaction term,
 it  describes an atomic Josephson  tunnelling process.
  $V_i({\bf x}) (i=1,2)$ are trapping potentials.
 Interactions between atoms are described by a nonlinear  self-interaction term
 $U_i=4\pi\hbar^2a^{sc}_i/m$ and a term that corresponds the nonlinear interaction
 between different component $U_{12}=4\pi\hbar^2a^{sc}_{12}/m$, where    $a^{sc}_i$
 is $s$-wave scattering lengths of component $i$ and  $a^{sc}_{12}$ that between
 component 1 and 2. For simplicity, throughout this paper we set $\hbar=1$, and
 assume that $a^{sc}_1=a^{sc}_2=a^{sc}$, $V_1({\bf x})=V_2({\bf x})$.

For weak many-body interactions, i.e., for   small number of
condensate  atoms, it has been well known that the Hamiltonian (1)
can reduce to a two-mode Hamiltonian \cite{gor,kua}
\begin{eqnarray}
\label{e5} \hat{H}&=&\omega_0(\hat{a}^{\dagger}_1\hat{a}_1 +
\hat{a}^{\dagger}_2\hat{a}_2)
+  q(\hat{a}^{\dagger2}_1\hat{a}^2_1 + \hat{a}^{\dagger2}_2\hat{a}^2_2) \nonumber \\
&& +  g(\hat{a}^{\dagger}_1\hat{a}_2 + \hat{a}^{\dagger}_2\hat{a}_1) + 2\chi\hat{a}^{\dagger}_1\hat{a}_1
\hat{a}^{\dagger}_2\hat{a}_2,
\end{eqnarray}
where $\hat{a}_i$    ($\hat{a}^+_i$) is correspondent-mode annihilation (creation)
 operators  with  $[\hat{a}_i, \hat{a}^{\dagger}_i]=1$,
 and  $q$, $\chi$ and $g$ are coupling constants which characterize the
strength of nonlinear and linear  interactions, respectively.

In general, the Hamiltonian (\ref{e5})can not be  exactly solved
except the two limiting cases (i) $g=0$, $q\neq 0$ and/or
$\chi\neq 0$; (ii) $g\neq 0$ and $q=\chi=0$.
 However, for the general case of $g\neq 0$, $q\neq 0$ and
$\chi\neq 0$, an approximate analytic solution of the Hamiltonian
(\ref{e5}) can be obtained by introducing  a  pair of  bosonic
operators: $\hat{A}_1=e^{-igt}(\hat{a}_1+\hat{a}_2)/\sqrt2$ and
$\hat{A}_2=ie^{igt}(\hat{a}_1 - \hat{a}_2)/\sqrt2$, which satisfy
the usual bosonic commutation relation: $[\hat{A}_i,
\hat{A}^{\dagger}_j]=\delta_{ij}$.
 Assume that the BEC is  initially in the coherent state defined
 by
\begin{equation}
\label{e6}
 |\alpha_1,\alpha_2\rangle
=\exp\left[-frac{1}{2}(|\alpha_1|^2+|\alpha_1|^2)\right]\sum^{\infty}_{n,m=0}\frac{\alpha^n_1\alpha^m_2}{\sqrt{n!m!}}|n,m\rangle,
\end{equation}
Where $\alpha_1$ and $\alpha_2$ are two arbitrary complex numbers.
Under the rotating wave approximation, the energy and wave
function of the system in the $(\hat{A}_1,\hat{A}_2)$
representation with the basis $\{|n,m)\}$ can be given  by
\begin{eqnarray}
\label{e7}
E(n,m)&=&\omega(n+m) + g(n-m) + \frac{1}{4}(3q+2\chi)\nonumber \\
&&\times (n+m)^2- \frac{q}{4}(n-m)^2 - \chi nm,\\
\label{e8} |\Phi(t)\rangle
&=&e^{-\frac{1}{2}N}\sum^{\infty}_{n,m=0}\frac{1}{\sqrt{n!m!}}
u_1^n(iu_2)^m \nonumber \\
&&\times e^{-iE(n,m)t}|n,m),
\end{eqnarray}
where  $n$ and $m$ take nonnegative integers, $\omega=\omega_0-(\chi-q)/2$,
 $N=|\alpha_1|^2+|\alpha_2|^2=|u_1|^2+|u_2|^2$ is the total number of atoms
 with two new parameters defined by
\begin{equation}
\label{e9}
u_1=\frac{1}{\sqrt2}(\alpha_1+\alpha_2), \hspace{0.5cm}
u_2=\frac{1}{\sqrt2}(\alpha_1-\alpha_2).
\end{equation}

\section{BEC Polarization states}

In order to discuss  the polarization properties  of the two-component BEC,
it is convenient to define the Hermitian Stokes operators \cite{col} as
\begin{eqnarray}
\label{e10}
\hat{S}_0&=&(\hat{a}^+_1\hat{a}_1+\hat{a}^+_2\hat{a}_2),
\hspace{0.3cm}
\hat{S}_1=(\hat{a}^+_1\hat{a}_1-\hat{a}^+_2\hat{a}_2), \nonumber \\
\hat{S}_2&=&(\hat{a}^+_1\hat{a}_2+\hat{a}^+_2\hat{a}_1), \hspace{0.3cm}
\hat{S}_3=i(\hat{a}^+_2\hat{a}_1-\hat{a}^+_1\hat{a}_2),
\end{eqnarray}
which satisfy the commutation relations:
$[\hat{S}_j, \hat{S}_k]=2i\epsilon _{jkl}\hat{S}_l$ and $[\hat{S}_j, \hat{S}_0]=0$.
 The noncommutability of the Stokes operators $S_1$, $S_2$,
and $S_3$ precludes the simultaneous measurement of the physical
quantities represented by them.

\begin{figure}
\begin{center}
\includegraphics[width=3.8in,height=3.6in]{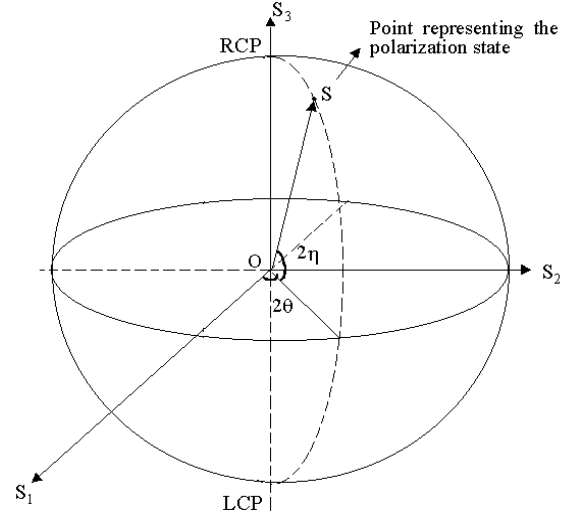}
\end{center}
\caption{Representation of the BEC polarization state  on the
Poincar\'{e} sphere.}
\end{figure}

These Stokes operators are quantum analog of the classical Stokes
parameters for a light field, i.e., $S_{\mu}=\langle
\hat{S}_{\mu}\rangle$, where $\mu=0,...,3$. Any polarization state
of the  BEC can be completely described by the four Stokes
parameters  which can be written as a string: $\{S_0, S_1, S_2,
S_3\}$. The first  parameter expresses the total number of atoms
in the BEC. The remaining three parameters describes the
polarization state.

For a completely unpolarized BEC only the parameter $S_0$ is
nonzero:
\begin{equation}
S_{\mu}=S_0\{1; 0; 0; 0\}.
\end{equation}

 The BEC is called completely polarized if the Stokes parameters  satisfy the
following relation:
\begin{equation}
S_0=(S^2_1+S^2_2+S^2_3)^{1/2}.
\end{equation}

A partially polarized BEC is a mixture of a completely unpolarized
BEC and  a completely polarized  BEC.   All of the four Stokes
parameters are independent, however, they satisfy the following
inequality $S_0>(S^2_1+S^2_2+S^2_3)^{1/2}$. And the degree of
polarization can be  measured by
\begin{equation}
\label{e11}
 P=\frac{\sqrt{S^2_1+S^2_2+S^2_3}}{S_0}.
\end{equation}
For a partially polarized BEC $0<P<1$, a completely polarized BEC has $P=1$,
while a completely unpolarized BEC has $P=0$.

BEC polarization states can be described by a Poincar\'{e} sphere
plotted in Fig. 1.  A particular polarization state of the BEC may
be represented as a point in a three-dimensional Stokes space by
introducing the Stokes vector  ${\bf S}$ with Cartesian
coordinates $S_1$, $S_2$ and $S_3$.  The point lies on a
Poincar\'{e} sphere  \cite{col} with  radius
 $(S^2_1+S^2_2+S^2_3)^{1/2}$. If the Stokes parameters are known,
the polarization azimuth $\Theta$ and the ellipticity angle $\eta$
defined by the following expressions,  respectively,
\begin{eqnarray}
\label{e12a}
\tan2\Theta&=&\frac{S_2}{S_1}, \\
\label{e12b} \sin2\eta&=&\frac{S_3}{S_0}.
\end{eqnarray}
 If the total number of the atoms, and therefore $S_0$, the degree of
polarization $P$, the polarization azimuth $\Theta$ and the
ellipticity $\eta$  of the polarization ellipse are known,  one
may return the Stokes parameters $S_{1-3}$:
\begin{eqnarray}
\label{e12}
S_1&=&PS_0\cos(2\eta)\cos(2\Theta), \nonumber \\
S_2&=&PS_0\cos(2\eta)\sin(2\Theta), \nonumber \\
S_3&=&PS_0\sin(2\eta).
\end{eqnarray}

If the BEC is totally polarized, the radius of the Poincar\'{e}
sphere equals $S_0$. The `north' and `south' poles of the sphere
correspond to right and left circular polarizations, respectively.
Points on the `equator' represent linear polarization with
`longitude' being twice the polarization azimuth $\Theta$. The
`latitude' of a point will give twice the angle of ellipticity
$\eta$. Antipodes, i.e., points at opposite ends
 of a diameter of the sphere, are  known as orthogonal polarizations.
If the BEC is partially polarized the ratio between the
Poincar\'{e} sphere radius and the Stokes parameter $S_0$ gives
the degree of polarization (\ref{e11}).

We now examine the dynamical evolution of polarization of the BEC.
From Eqs.(\ref{e7}) and (\ref{e8}) we can calculate
 the Stokes parameters. The results are
 \begin{eqnarray}
\label{e13}
 S_0&=&N, \\
\label{e14}
S_1&=&2|u_1||u_2|\cos[4gt+\theta(t)] \nonumber \\
   & &\times \exp[-2N\sin^2(\frac{q-\chi}{2})t],  \\
\label{e15}
S_2&=&2|\alpha_1||\alpha_2|\cos(\varphi_{\alpha_1}-\varphi_{\alpha_2}),  \\
\label{e16}
S_3&=&-2|u_1||u_2|\sin[4gt+\theta(t)] \nonumber \\
   & &\times \exp[-2N\sin^2(\frac{q-\chi}{2})t].
\end{eqnarray}
where we have set $u_i=|u_i|e^{i\varphi_{u_i}}$ and the   symbol:
\begin{equation}
\label{e17}
 \theta(t)=
(\varphi_{u_2}-\varphi_{u_1})+(|u_2|^2-|u_1|^2)\sin(q-\chi)t,
\end{equation}

Assume that the system is initially in  a state  $|\alpha_1,0>$ of
$\alpha_1$ being real and $\alpha_2=0$, which implies that the BEC
is initially in a linear  horizontal polarization state with the
Stokes parameters given by  $S_{\mu}=S_0\{1; 1; 0; 0\}$. From
Eqs.(\ref{e13}) to (\ref{e15}), one can obtain some interesting
polarization states of the BEC at some specific times. For
instance, at times $t_k=(2k+1)\pi/(8g)$ with $k$ being an integer,
the BEC is in a left-hand circular polarization state with the
Stokes parameters being $S_{\mu}=S_0\{1; 0; 0; -1\}$. When
$t_k=(2k-1)\pi/(8g)$ with $k$ being an integer, the BEC is in  a
right-hand circular polarization state with the Stokes parameters
being $S_{\mu}=S_0\{1; 0; 0; 1\}$. At times $t_k= k\pi/(2g)$ with
$k$ being an integer, the BEC returns the initial polarization
state, a linear  horizontal polarization state. When
$t_k=(2k+1)\pi/(4g)$ with   $k$ being an integer, one may find
that the BEC is in  a linearly vertical polarization state with
the Stokes parameters given by $S_{\mu}=S_0\{1; -1; 0; 0\}$.

\begin{figure}
\begin{center}
\includegraphics[width=3.8in,height=3.6in]{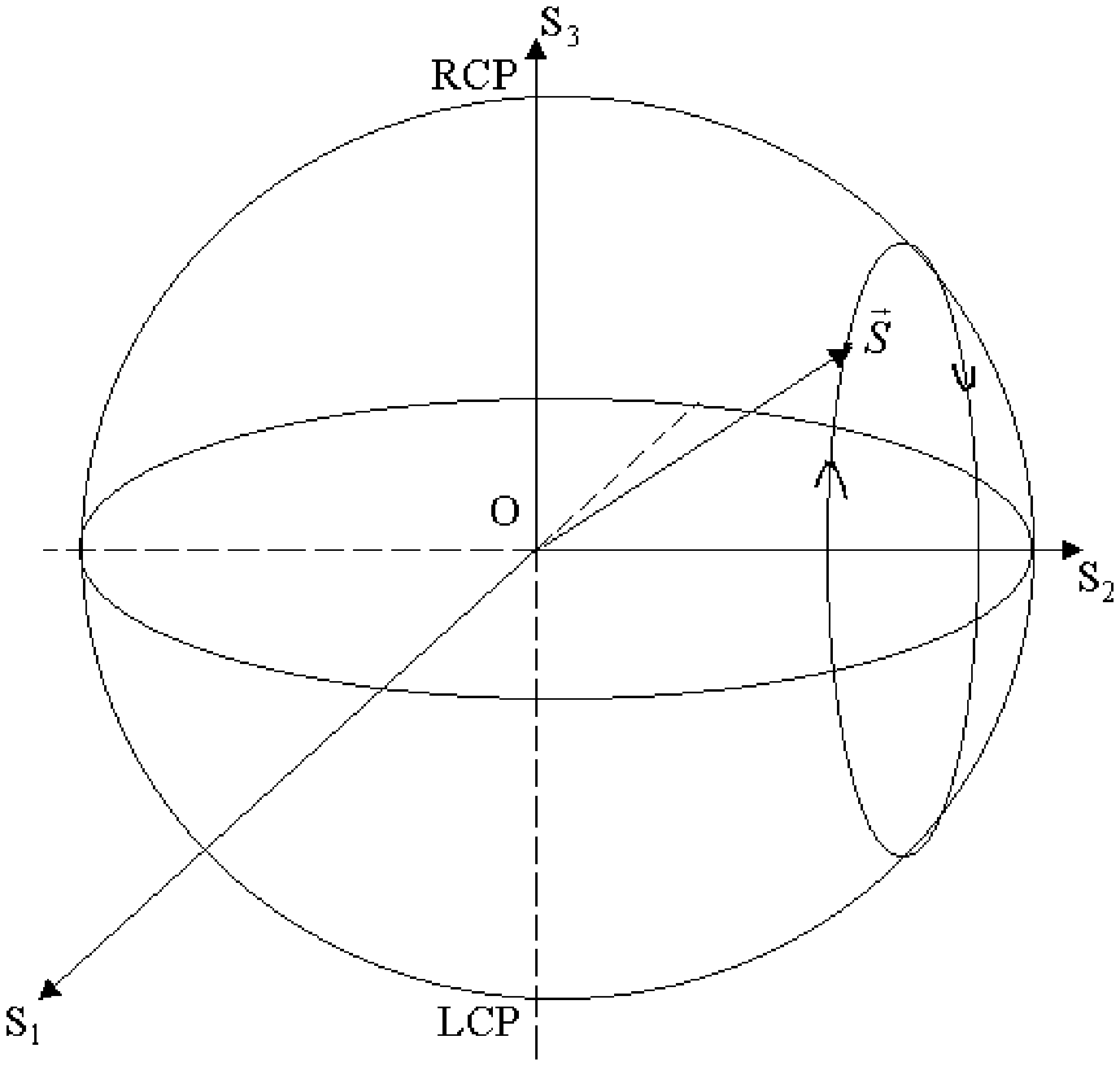}
\end{center}
\caption{Precession of the Stokes vector  on the Poincar\'{e}
sphere in the BEC evolution when $q=\chi$ and without
decoherence.}
\end{figure}


 From Eqs. (11) to (14), we can obtain the degree of polarization
\begin{eqnarray}
\label{e18}
 P(t)&=&\frac{1}{N}\left\{(|u_2|^2-|u_1|^2)^2 + 4|u_1|^2|u_2|^2\right. \nonumber\\
&&\times\left.\exp\left[-4N\sin^2\left(\frac{q-\chi}{2}\right)t\right]\right\}^{1/2}.
\end{eqnarray}

From above equation we see that: (i) the BEC system exhibits
periodic polarization oscillations with a period $T=4\pi/|q-\chi|$
which depends on the difference between the self-interaction and
inter-component interaction. (ii) the degree of polarization
depends only nonlinear interactions and the initial state of the
BEC but not linear interaction due to the independence of $P(t)$
upon the linear coupling. (iii) When $q=\chi$, we obtain
$P(t)=P(0)=1$, which means that  the degree of polarization
remains unchanged. It is interesting to note that  when $q=\chi$,
from Eqs. (\ref{e13}) to (\ref{e15}), we can find that the
evolution equation of the Stokes vector takes the following form
\begin{equation}
\label{e19}
 \frac{d{\bf S}}{dt}=-{\bf \Omega}\times{\bf S},
\end{equation}
where ${\bf \Omega}=4g\hat{e}_2$ with $\hat{e}_2$ being the unit
vector of the $S_2$ axis. Geometrically, Eq.(\ref{e19}) indicates
that the Stokes vector ${\bf S}$  precesses around the $S_2$ axis
in the dynamic evolution of the BEC. The value of the processing
frequency $\Omega$ is determined by the strength of the BEC linear
coupling. In Fig.2 we plot the precession of the Stokes vector in
the BEC evolution.

\section{Influence of  decoherence on polarization}
We now consider the effect of the decoherence on the BEC
polarization. As is well known, any  system inevitably interacts
with  its environment. In fact, in current experiments on BECs of
dilute alkali atomic gases condensate   atoms continuously
interact with non-condensate atoms (reservoir). Interaction
between a BEC and environment causes  decoherence \cite{kua,ang}.
 We use a reservoir consisting of an infinite set of
harmonic oscillators to model environment of condensate atoms in a
trap, and assume the total Hamiltonian \cite{kua} to be
\begin{eqnarray}
\label{e20} \hat{H}_T&=&\hat{H} +
\sum_k\omega_k\hat{b}^{\dagger}_k\hat{b}_k +
F(\{\hat{O}\})\sum_kc_k(\hat{b}^{\dagger}_k+\hat{b}_k)
\nonumber \\
&&+F(\{\hat{O}\})^2\sum_k\frac{c_k^2}{\omega_k^2},
\end{eqnarray}
where the second term is the Hamiltonian of the reservoir. The
last term in Eq.(\ref{e20}) is a renormalization term \cite{aoc}.
The third term in Eq.(\ref{e20})  represents the interaction
between the system and the reservoir with a coupling constant
$c_k$, where $\{\hat{O}\}$ is a set of linear operators of the
system or their linear combinations in the same picture as that of
$\hat{H}$, $F(\{\hat{O}\})$ is an operator function  of
$\{\hat{O}\}$. In order to  enable
 what the interaction between the system and  environment describes is decoherence
 not dissipation \cite{sun},
 we require that the linear operator $\hat{O}$  commutes with the Hamiltonian
 of the system $\hat{H}$.   The concrete form of the function $F(\{\hat{O}\})$,
 which may be considered  as  an experimentally determined quantity. Therefore, the decohering interaction
in (\ref{e20})  can not only describe decoherence caused by  the
effect of elastic collisions between condensate and non-condensate
atoms, but also simulate decoherence caused by other decohering
sources \cite{klm} through properly  choosing  the operator
function of the system $F(\{\hat{O}\})$.

 From Hamiltonian (\ref{e20}) through  lengthy calculations  we can get the Stokes parameters
\begin{eqnarray}
\label{e21}
S_0(t)&=&N, \\
\label{e22}
S_1(t)&=&2\sum_{m,n}\sin'\phi'_{mn}(t)\exp[-\gamma_{(m,n)(m+1,n-1)}(t)], \\
\label{e23}
S_2(t)&=& \sum_{m,n} \cos''\phi''_{mn}(t)\exp[-\gamma_{(m,n)(m,n)}(t)], \\
\label{e24}
S_3(t)&=&\sum_{m,n}\cos'\phi'_{mn}(t)\exp[-\gamma_{(m,n)(m+1,n-1)}(t)] \nonumber  \\
      & &-\sum_{m,n} \sin''\phi''_{mn}(t) \exp[-\gamma_{(m,n)(m,n)}(t)].
\end{eqnarray}
where we have introduced symbols
\begin{eqnarray}
\label{e25}
\sin'\phi'_{mn}(t)&=&\sqrt{n(m+1)}|\rho_{(m,n)(m+1,n-1)}(0)|  \nonumber \\
                  & &\times\sin[2gt+\phi_{(m,n)(m+1,n-1)}(t)],  \\
\label{e26}
\cos'\phi'_{mn}(t)&=&\sqrt{n(m+1)}|\rho_{(m,n)(m+1,n-1)}(0)|  \nonumber \\
                  & &\times\cos[2gt+\phi_{(m,n)(m+1,n-1)}(t)],  \\
\label{e27}
\sin''\phi''_{mn}(t)&=&(m-n)|\rho_{(m,n)(m,n)}(0)|  \nonumber \\
      & &\times \sin[2gt+\phi_{(m,n)(m,n)}(t)],\\
\label{e28}
\cos''\phi''_{mn}(t)&=&(m-n)|\rho_{(m,n)(m,n)}(0)|   \nonumber \\
      & &\times \cos[2gt+\phi_{(m,n)(m,n)}(t)],
\end{eqnarray}
where the damping factor and the phase shift induced by interaction between the
BEC and reservoir are given  by
 \begin{eqnarray}
 \label{e29}
\gamma_{(m',n')(m,n)}(t)&=&v^2_-(m',n';m,n)Q_2(t), \\
\label{e30}
\phi_{(m',n')(m,n)}(t)&=&v_+(m',n';m,n)v_-(m',n';m,n)Q_1(t)\nonumber  \\
                 & & +\theta_{(m',n')(m,n)},
\end{eqnarray}
Here we have introduced the  notations
 \begin{eqnarray}
 \label{e31}
  v_{\pm}(n,m;n',m')&=&F(\{O(n,m)\}) \pm F(\{O(n',m')\}), \\
\label{e32}
\rho_{(m',n')(m,n)}(0)&=&|\rho_{(m',n')(m,n)}(0)\nonumber \\
&&\times |\exp[-i\theta_{(m',n')(m,n)}].
\end{eqnarray}
where the two reservoir-dependent functions are given by
\begin{eqnarray}
\label{e33}
 Q_1(t)&=&\int^{\infty}_{0} d\omega
J(\omega)\frac{c^2(\omega)}{\omega^2}\sin(\omega t), \\
\label{e34}
 Q_2(t)&=&2\int^{\infty}_{0} d\omega J(\omega)
\frac{c^2(\omega)}{\omega^2}\nonumber\\
&&\times\sin^2\left (\frac{\omega t}{2}\right)\coth\left
(\frac{\beta\omega}{2}\right ),
\end{eqnarray}
where $J(\omega)$ is the spectral density of the reservoir,
$c(\omega)$ is the   continuum expression for $c_k$, and
$\beta=1/k_BT$ with $k_B$ and $T$ being  the Boltzmann constant
and temperature, respectively.

It is well known that decoherence corresponds to the decay of
off-diagonal  elements of the reduced density matrix of a quantum
system. For the case under our consideration, the degree of
decoherence is determined by the decaying factor given by
Eq.(\ref{e29}) .  It is interesting  to note that if we choose a
proper operator function $F(\{\hat{O}\})$  to make
$F(\{O(m',n')\})=F(\{O(m,n)\})$ for
 $(m',n')\neq (m,n)$, then we find that  $\rho_{(m',n')(m,n)}(t)=\rho_{(m',n')(m,n)}(0)$
which indicates that the decoherence-free  evolution of the BEC is
realized,  and the BEC preserves its initial polarization state.
From Eq. (\ref{e22})  into (\ref{e24})   we can immediately draw
one important qualitative conclusion:  since $Q_2(t)$ is positive,
the existence of the decoherence is always to tend to make  Stokes
parameters decay. Therefore, decoherence always depolarizes the
BEC.

From Eqs.(\ref{e22})  to (\ref{e24}) , (\ref{e29})  and
(\ref{e30}) we see that all necessary information
 about the effects of the reservoir on the polarization is
contained in the spectral density of the reservoir. To proceed
further let us now specialize to the Ohmic  case \cite{cha} with
the  spectral distribution
\begin{equation}
\label{e35}
J(\omega)=\frac{\kappa\omega}{c^2(\omega)}\exp\left(-\frac{\omega}{\omega_c}\right),
\end{equation}
 where $\omega_c$ is the high frequency cut-off, $\kappa$
is a positive characteristic parameter of the reservoir. With this
choice,  when $\omega_ct\gg 1$,  at zero temperature
 we have $Q_2(t) \doteq \kappa\ln(\omega_ct)$, so that the components of the Stokes
 vector are given by
\begin{eqnarray}
\label{e36}
S_1(t)&=&2\sum_{m,n}\sin'\phi'_{mn}(t)(\omega_ct)^{-\kappa v^2_-(m,n;m+1,n-1)}, \\
\label{e37}
S_2(t)&=& \sum_{m,n} \cos''\phi''_{mn}(t) (\omega_ct)^{-\kappa v^2_-(m,n;m,n)}, \\
\label{e38}
S_3(t)&=&-\sum_{m,n} \sin''\phi''_{mn}(t) (\omega_ct)^{-\kappa v^2_-(m,n;m,n)}  \nonumber \\
      & &+ \sum_{m,n}\cos'\phi'_{mn}(t)(\omega_ct)^{-\kappa v^2_-(m,n;m+1,n-1)}.
\end{eqnarray}
which indicate that Stokes parameters  decay according to the
`power law'. Therefore, geometrically decoherence makes the Stokes
vector contract towards the center of the Poincar\'{e} sphere
according to the `power law'  with the time evolution.

At finite temperature, we have $Q_2(t)\doteq \kappa[\ln(\frac{\beta\omega_c}{2\pi})+ \frac{\pi t}{\beta}]$, so that
\begin{eqnarray}
\label{e39}
S_1(t)&=&2\sum_{m,n}\sin'\phi'_{mn}(t)
      (\frac{\beta\omega_c}{2\pi})^{-\kappa v^2_-(m,n;m+1,n-1)}\nonumber \\
      & &\times \exp[-\kappa\pi\beta^{-1}v^2_-{(m,n;m+1,n-1)}t], \\
\label{e40}
S_2(t)&=& \sum_{m,n} \cos''\phi''_{mn}(t)
      (\frac{\beta\omega_c}{2\pi})^{-\kappa v^2_-(m,n;m,n)} \nonumber \\
      & &\times \exp[-\kappa\pi\beta^{-1}v^2_-{(m,n;m,n)}t], \\
\label{e41}
S_3(t)&=&\sum_{m,n}\cos'\phi'_{mn}(t)
      (\frac{\beta\omega_c}{2\pi})^{-\kappa v^2_-(m,n;m+1,n-1)} \nonumber \\
      & &\times \exp[-\kappa\pi\beta^{-1}v^2_-(m,n;m+1,n-1)t], \nonumber \\
      & &-\sum_{m,n} \sin''\phi''_{mn}(t)
      (\frac{\beta\omega_c}{2\pi})^{-\kappa v^2_-(m,n;m,n)}\nonumber \\
      & &\times \exp[-\kappa\pi\beta^{-1}v^2_-(m,n;m,n)t],
\end{eqnarray}
which indicate that at finite temperature  Stokes parameter   decay
  according to the  `exponential law'. Geometrically this means that decoherence
  makes the Stokes vector contracts toward the center of the Poincar\'{e} sphere
    according to  the `exponential law'  with the time evolution. Therefore,
 decoherence depolarizes the BEC.

 From Eqs. (\ref{e36}) to (\ref{e41}) we see that a BEC initially in a completely  polarized state
 ($P=1$) evolves into a completely unpolarized state ($P=0$) when the time approaches the infinity
 due to the influence of decoherence.

\section{Concluding remarks}
In conclusion, we have examined the nonlinear polarization
dynamics of a two-component BEC and influence of decoherence
induced by environment on BEC polarization through introducing
four BEC Stokes operators. We have shown that different
polarization states of the BEC may be geometrically represented on
a Poincar\'{e} sphere. Without decoherence, it has been found that
nonlinear inter-atomic interactions in the BEC induce periodic
polarization oscillations whose periods depend on the nonlinear
interaction strengths.  When nonlinear self-interaction ($q$)
equals inter-component nonlinear interaction ($\chi$), the BEC
initially in a completely  polarized state remains a completely
polarized state, however, the Stokes vector  precesses around the
$S_2$ axis with a precessing frequency determined by the linear
coupling strength. When decoherence is taken into account, we find
each component of the Stokes vector decays. This indicates that
decoherence depolarizes the BEC. In particular, We have obtained
analytical expressions of the Stokes parameters and found  that
for the reservoir-spectral density of the  Ohmic case, the stokes
parameters  decay  by the `power law' at zero temperature, and the
`exponential law' at finite temperature, respectively.

 \begin{center}
 {ACKNOWLEDGMENTS}
 \end{center}
 L.M.K.  acknowledges support from  the National ``973" Research Plan,
the National Natural Science Foundation,  EYTF of the Educational
Department of China, and Hunan Province STF. This work was also
supported in part by grants from Hong Kong Research Grants Council
(RGC) and the Hong Kong Baptist University Faculty Research Grant
(FRG).

\end{document}